\documentclass[preprint, superscriptaddress, showkeys]{revtex4-2}

\usepackage{hyperref}
\hypersetup{colorlinks=true, citecolor=blue, urlcolor=blue, linkcolor=blue}

\usepackage{lineno}
\usepackage{graphicx}
\usepackage{amsmath}
\usepackage{cases}
\usepackage{booktabs}
\usepackage{chemformula} 
\usepackage[T1]{fontenc} 
 \usepackage{mathtools, amssymb, amsmath}
 \usepackage{xcolor}
\usepackage{hyperref}

 \newcommand{\mpq}{Université Paris Cité, Laboratoire Matériaux et Phénomènes Quantiques,  75013 Paris, France}
 \newcommand{\ctwon}{Université Paris-Saclay, CNRS, Centre for Nanosciences and Nanotechnology, 91120 Palaiseau, France}
 \newcommand{\diet}{Sapienza University of Rome, Dipartimento di ingegneria elettronica e telecomunicazioni, 00184 Roma, Italy}
 \newcommand{\cnr}{CNR-INO, Istituto Nazionale di Ottica, 80078 Pozzuoli, Italy}
 \newcommand{\TRT}{Thales Research and Technology, 91120 Palaiseau, France}
\newcommand{\IUF}{Institut universitaire de France, France}

\begin{document}

\title{Dispersion engineered AlGaAs-on-insulator nanophotonics by distributed feedback}

\author{Francesco Rinaldo Talenti}
\affiliation{\ctwon}

\author{Luca Lovisolo}
\affiliation{\mpq}
\affiliation{\ctwon}

\author{Zijun Xiao}
\affiliation{\ctwon}

\author{Zeina Saleh}
\affiliation{\ctwon}

\author{Andrea Gerini}
\affiliation{\mpq }

\author{Carlos Alonso-Ramos}
\affiliation{\ctwon}

\author{Martina Morassi}
\affiliation{\ctwon}

\author{Aristide Lemaître}
\affiliation{\ctwon}

\author{Stefan Wabnitz}
\affiliation{\diet}
\affiliation{\cnr}

\author{Alfredo De Rossi}
\affiliation{\TRT}

\author{Giuseppe Leo}
\affiliation{\mpq}
\affiliation{\IUF}

\author{Laurent Vivien}
\affiliation{\ctwon}


\begin{abstract}
Technological advances in the fabrication of nanophotonic circuits have driven the scientific community to increasingly focus on the precise tailoring of their key optical properties, over a broadband spectral domain. In this context, the modulation of the local refractive index can be exploited to customize an effective reflectivity by the use of distributed Bragg mirrors, enabling the on-chip integration of Fabry-Pérot resonators. The resulting cavity length is strongly wavelength-dependent, offering practical solutions to the growing demand of dispersion engineering. Owing to their typically high core-to-cladding refractive index contrast and exceptional nonlinear properties, III-V semiconductor-based platforms represent promising candidates for the fabrication of Bragg reflectors. In this work, we propose an AlGaAs-on-insulator linear resonator based on distributed Bragg mirrors. We discuss the first experimental demonstration of a systematic, shape-constrained inverse design technique which tailors a prescribed dispersion profile, showing a strong agreement between simulations and measurements. In perspective, the proposed approach offers an efficient and general response to the challenge of dispersion engineering in integrated optical circuits.
\end{abstract}

\maketitle

\section{Introduction}
\label{sec:intro}

Fabry–Pérot (FP) resonators have recently been developed on fiber-based platforms (Fiber Fabry–Pérot, FFP), resulting in compact, efficient, and relatively easy to implement systems for the demonstration of modulation instability (MI) \cite{Bunel_2023}, frequency combs and cavity soliton generation \cite{Bunel_2024, Bourcier_2024}. The on-chip integration of such systems--driven by the goal of minimizing the device footprint and reducing the power operation--requires the use of distributed Bragg reflectors (DBRs). First prototypes of this sort have been proposed for Kerr comb generation \cite{Xu_2016} and then implemented on silicon nitride (SiN) \cite{Yu_2019} to enhance four-wave-mixing (FWM) \cite{Xie_2020}, or to trigger solitons formation \cite{Wildi_2023}, while parent demonstrations are those of Moiré plasmonic cavities \cite{Kocabas_2009,Tharrault_2025} suitable for laser applications \cite{Karademir_2015}. Similar systems have been also conceived with III-V semiconductor, and proven to be promising for nonlinear frequency conversion \cite{Ye_2025}, pulse spectral broadening \cite{Nardi_2024} and the design of nanophotonic distributed feedback (DFB) lasers \cite{Bourgon_2025, Bourgon_2025_preprint}.
\\

In a DFB structure, a periodic longitudinal modulation of the refractive index is exploited to introduce a Bragg backward scattering mechanism \cite{Kogelnik_1971,Shank_1971}. As a result, two counter-propagating optical waves are spatially confined and linearly coupled by multiple local reflections along the cavity\cite{Kogelnik_1972}. DFB systems, thus, inhibit light propagation within a finite spectral domain \cite{10.1063/1.1653605,Yablonovitch_1987}. In strict analogy and with a direct connection with electronic dispersion diagrams, photonic band structures can be drawn for dielectrics which are geometrically organized with a given and well defined periodicity $\Lambda=\lambda_0/2n$, where $\omega_0/2\pi= c/\lambda_0$ is the central frequency of oscillation (i.e. the Bragg frequency), $c$ is the speed of light in vacuum and $n$ is the refractive index of the bulk material\cite{Yablonovitch_1993}.\\

The underlying physics is that of a photonic crystal (PhC), whose geometry determines the effective local band diagrams. Furthermore, the introduction of a crystal defect induces light  confinement, and it can be exploited for the design of nano-cavities. The maximization of the quality factors, though, is a non-trivial task.
First convincing demonstrations of low-dimensional (i.e. $\le$2D) PhC cavities introduced a novel light confining mechanism, named \textit{gentle confinement} \cite{Akahane_2003, Akahane_2005}. The  basic idea consists in the fact that light scattering at defect boundaries can be a detrimental source of loss. To avoid that, light must be progressively (or gently) guided towards the confining spatial domain, by adiabatically tuning the geometry of the PhC unit cells surrounding the defect. To use the words of the inventors of this mechanism, \textit{light should be confined gently in order to be confined strongly} \cite{Akahane_2003}. The implementation of this simple, but yet paradigmatic objective is a primordial dispersion engineering technique, based on tuning the PhC geometry around the defect. Following these ideas, the design of a bichromatic potential involving the superimposition of two 1D lattices with quasi-identical periods \cite{Harper_1955,aubry_1980}, has been proposed for PhC slabs \cite{Alpeggiani_2015}, and proven to be highly efficient \cite{Simbula_2017,Asano_2017}. The resulting cold-cavity spectrum forms a comb of Hermite–Gauss modes confined by a parabolic potential, resembling the behavior of a quantum harmonic oscillator \cite{Combrie_2017}. The extremely low pump power which is required to trigger parametric oscillation indicates that these systems are particularly promising for quantum and nonlinear optics \cite{Marty_2020}. Inverse design approaches can be applied to engineer the dispersion \cite{Minkov_2014, Talenti_2022}, to maximize the quality factors \cite{Minkov_2020} or to minimize losses in waveguide configurations \cite{Wang_2012}.\\


Among the different platforms of interest, III-V semiconductors, and specifically AlGaAs and AlGaAs-on-insulator (AlGaAs-OI), are very promising for the fabrication of DFB structures for nonlinear optics applications, considering the large $\chi^{(2)}$ and $\chi^{(3)}$ response of such materials, combined with the possibility of suppressing two-photon absorption (TPA) processes at telecom wavelengths. 
The high core-to-cladding refractive index contrast, compared to that of more mature technologies such as SiN-on-insulator (SiN-OI)\cite{Yu_2019,Wildi_2023}, suggests the potential for broadband tunability of the chromatic dispersion profile. This is allowed by the high degree of flexibility in tuning the local refractive index, which also ensures strong optical confinement and compatibility with the Silicon photonics technology \cite{Jiang_2020}. AlGaAs provides a suitable solution for the efficient integration of parametric microcomb sources \cite{Pu_2016}, with record low-power thresholds \cite{Chang_2020} and the perspective for opening a $\omega\rightleftarrows 2\omega$ bandwidth via second-harmonic generation processes \cite{Talenti_2025}. \\

In this work, we carry out the design of an AlGaAs-OI 1D cavity comprised by two identical DBRs \cite{Ye_2025, Talenti_2025_conf_DBR} which are optimized for dispersion management purposes. The shaping of the PhC unit cells composing the mirrors turns out to be a very efficient tool for tuning the local badgap and the resulting effective reflectivity. Next, we report what is, to our knowledge, a first experimental demonstration of a recently proposed shaped-constrained inverse design technique \cite{Talenti_2022}, which permits to tailor a prescribed dispersion profile by tuning the local reflectivity of a multimode PhC nanobeam resonator.\\

The manuscript is organized as follows: in section~\ref{sec:methods} we present the basic principles of the design technique and its associated methods; in section \ref{sec:PhC cavity} we discuss the implementation of a PhC cavity, whose unit cell can be exploited in order to customize the device reflection; in section \ref{sec:dispersion tailored FP-PhC} we report the properties of our dispersion engineered devices and their experimental characterization; in section \ref{sec:Conclusions} we draw our conclusions and perspectives.

\section{DBR design for efficient bandgap tuning} 
\label{sec:methods}

\begin{figure}[!b]

  \includegraphics[width=\textwidth]{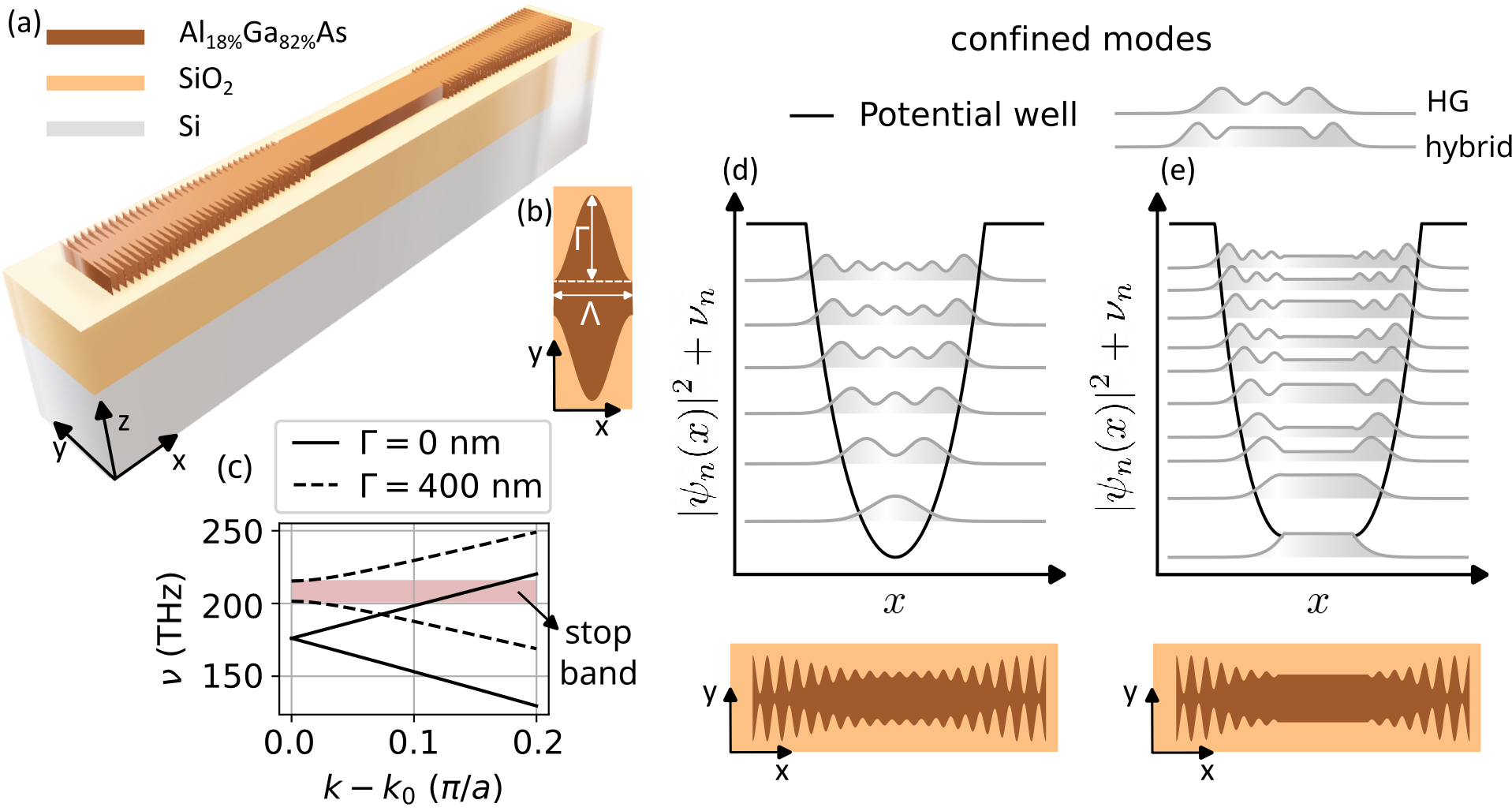}
  \caption{(a) Schematic of the device. (b) PhC unit cell. (c) Photonic band diagram for specified PhC waveguide geometries. Spatial confinement of a PhC cavity (d) and of a FP-PhC (e). The resulting eignemodes have HG shaped or hybrid profiles, respectively. }
  \label{fig1:scheme}

\end{figure}

In Figure \ref{fig1:scheme} we present the schematic of our design. The device (a) consists of a (100) 400 nm thick Al$_{18\%}$Ga$_{82\%}$As on a 2 µm thick buried oxide layer, with a Si substrate. The cavity is comprised by a central waveguide section with symmetric DBRs placed at its sides. The unit cell of the mirrors is sketched in panel (b): it involves a sinusoidal corrugation of the waveguide stripe, with a fixed period $\Lambda=0.33$ $\mu$m and an amplitude $\Gamma/2$, which alters the local width of the waveguide and opens a photonic bandgap. The effective local waveguide width ($w_{\mathrm{eff}}$) along the propagation direction $x$ reads as:
\begin{equation}
    w_{\mathrm{eff}}(x)=w_0 + \dfrac{\Gamma(x)}{2}\sin\left(\dfrac{2 \pi x}{\Lambda}\right) \ \ \ , 
    \label{eq:weff}
\end{equation}
where we fix the average width at $w_0=0.61\ \mu$m, in order to guarantee broadband nonlinear operation \cite{Talenti_2025}. Notably, with these prescriptions and by fixing all other parameters, we are able to directly connect the local band structure to the corrugation amplitude $\Gamma(x)$. This is sketched in Figure \ref{fig1:scheme} (c), where we illustrate the band diagrams relative to the cases with either $\Gamma=0$ or 400 nm for the TE$_{00}$ mode, respectively, calculated by means of a commercial Finite Difference Time Domain (FDTD) solver (Lumerical Inc.). While $\Gamma=0$ nm corresponds to the case of a standard waveguide--which sustains the propagation of an unperturbed traveling wave in the $k$-$\nu$ space--with a finite corrugation amplitude (i.e., $\Gamma=400$ nm) the dispersion is modified and we observe the emergence of a stop-band where propagation is forbidden. As consequence, when a wave hits the region covered by the DBRs, it is partially back reflected and linearly coupled into a counter-propagating wave \cite{Shank_1971, Kogelnik_1972}. 

Appropriately designed DBRs introduce an effective potential well, that spatially confines the cavity eigenmodes, in close analogy with the problem of an effective mass $m_{\mathrm{eff}}=\hbar^{-1}\partial^2\omega/\partial k^2$ confined by a parabolic potential well \cite{Sun_2019}. This approach is especially relevant for pure PhC cavity configurations, which are achievable by placing two identical mirrors face-to-face. This situation is reported in Figure \ref{fig1:scheme} (d): a parabolic potential well confines Hermite-Gauss (HG) eigenmodes along the longitudinal direction $x$. Thanks to the local reflections induced by the waveguide corrugation, such structure couples a wave propagating along the forward direction ($\mathcal{A}^{+}$) with a feedback signal ($\mathcal{A}^{-}$) \cite{Kogelnik_1972}. Mathematically, a system that couples the one-dimensional counter-propagation of two confined waves reads as:

\begin{equation}
    \left( \mathcal{D} \partial^2_x \pm i v_g \partial_x  +\omega_0-\omega\right)\mathcal{A}^{\pm}+\mathcal{K}\mathcal{A}^{\pm}=0 \ \ \ ,
    \label{eq:RM_spatial}
\end{equation}
where  $v_g$ and $\mathcal{D}$ are the group velocity and its dispersion, respectively. $\mathcal{K}$ is the coupling between the two waves, or the grating strength, which quantifies the local photonic bandgap and the effective reflectivity. We refer to the system of coupled Eqs.(\ref{eq:RM_spatial}) as the reduced model (RM), since it provides a robust approximation of the full complex 3D dynamics\cite{Talenti_2022}. The potential well $V(x)$ can be drawn by mapping the photonic band-edge, i.e.:
\begin{equation}
    V(x)= \omega_0(x) + \dfrac{1}{2}\mathcal{K}(x) \ \ \ .
    \label{eq:V_eff}
\end{equation}
If $V(x)$ is parabolic, i.e. $V(x)\propto x^2$, it turns out that the eigenmodes of Eqs.(\ref{eq:RM_spatial}) are close to HG-shaped functions with a spatial envelope $\psi_n(x)$ and eigenvalues $\omega_n=2\pi\nu_n$, as reported in Figure \ref{fig1:scheme}(d) in normalized units. It is worth mentioning that this picture is exact in the Gross-Pitaevskii limit $V(x)\rightarrow \propto m_{\mathrm{eff}}x^2$ \cite{Sun_2019}. The ideal HG mode structure can be approximated by a PhC cavity formed by two identical mirrors with a corrugation which increases adiabatically as $\Gamma(x)\propto x^2$.\\
Therefore, the design of an effective parabolic potential well requires
an optimization procedure, which is reminiscent of inverse design, for defining the modulation profile of $\Gamma(x)$. Although in some cases the HG modes can be approximated without a full optimization \cite{Alpeggiani_2015,Simbula_2017,Combrie_2017,Marty_2020}, to our knowledge a systematic approach to master the dispersion and the modes structure has not been experimentally attempted yet \cite{Chopin_2022}. 

Here we model the DBR feedback by means of a compact and well-established coupled waves theory \cite{Shank_1971, Kogelnik_1971, Kogelnik_1972}, which reduces the dimensionality of the associated eigenvalues problem from 3 to 1. A computational algorithm based on the RM of Eqs.(\ref{eq:RM_spatial}) turns out to be orders of magnitude faster than standard and commonly used solvers, such as finite element method (FEM) or FDTD, applied to the full 3D problem. Interestingly, subject to a careful and proper calibration of the model, with the RM one may recover the same accuracy of exact 3D solvers. Meaning that one can use a highly precise and rapid eigenvalue solver, which is suitable for implementing systematic design strategies in order to tailor prescribed optical properties. It is worth noting that computational performance plays a critical role in the implementation of optimization algorithms, since these typically require a few hundreds, at least, of function evaluations \cite{Talenti_2022}. 
\\
Among the various dispersion engineered nanophotonic devices that have attracted significant attention in recent years, let us consider linear resonators integrated on-chip \cite{Xu_2016,Yu_2019,Wildi_2023}. An example of a such structure, and of its associated optical confinement, is sketched in Figure \ref{fig1:scheme} (e): within a homogeneous central section comprised by two identical DBRs, the two counter-propagating waves $\mathcal{A}^{\pm}$ are no longer linearly coupled. The eigenmodes amplitude is a constant until $\mathcal{A}^+$ hits the mirrors. At this point, the forward-propagating wave is partially back-reflected and coupled to backward wave $\mathcal{A}^-$. The penetration depth through the mirroring region strongly depends on the carrier wavelength. We can see how the confined modes take an hybrid form: they are HG-shaped in the feedback section, similar to the PhC cavity case, whereas their amplitude is constant in the central region, as it occurs in a FP resonator. For this reason, such a cavity is often addressed to as FP-PhC or, similarly, as Fabry–Pérot Bragg grating (FPBG).

\section{The PhC nanobeam cavity}
\label{sec:PhC cavity}

\begin{figure}[!b]
  \includegraphics[width=\textwidth]{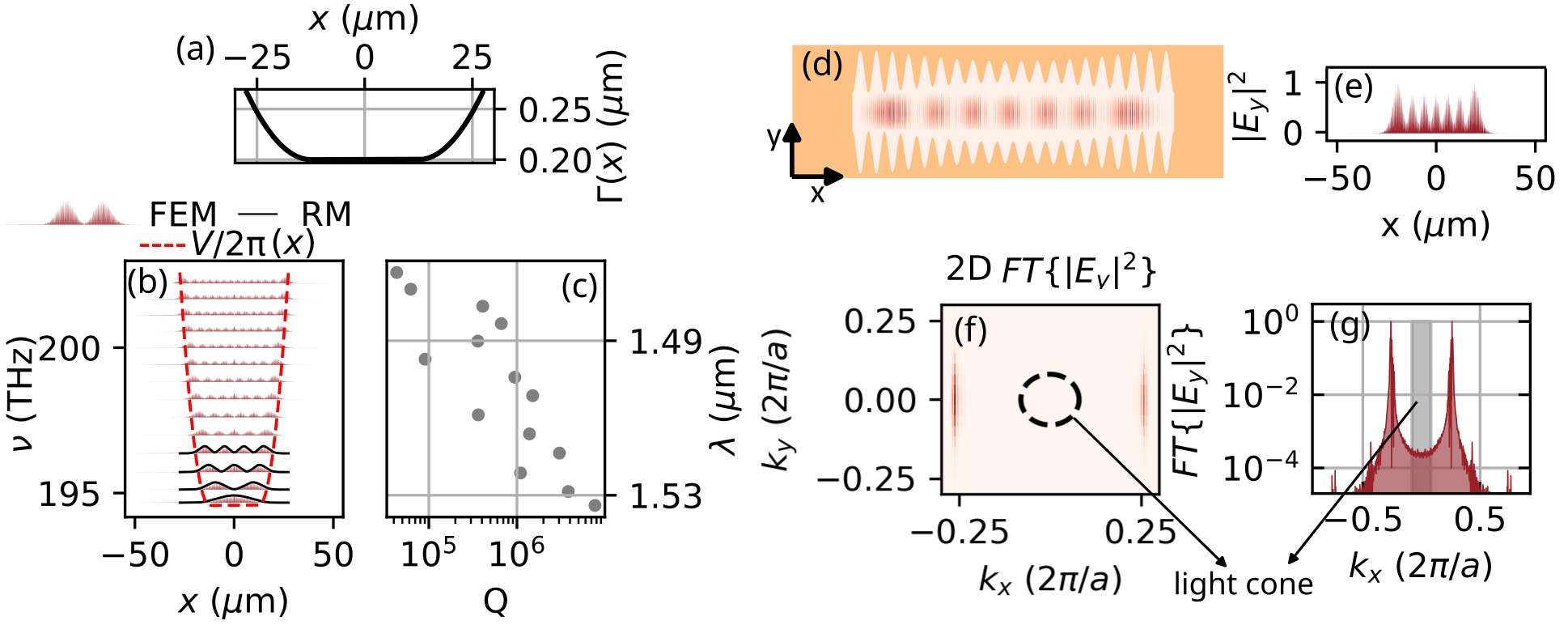}
  \caption{  (a) Profile of the quantity $\Gamma$ along the PhC cavity. (b) Cavity spectrum and confining potential well with the corresponding (c) intrinsic quality factor of the modes. (d) 2D and (e) 1D spatial optical confinement of an high-order HG mode, and (f,g) corresponding 2D and 1D Fourier transforms.}
  \label{fig2:PhC0_design}
\end{figure}

The dispersion engineering of a FP-PhC is strongly influenced by the design and fabrication quality of the Bragg mirrors. Thus, let us proceed to describe the fabrication and the linear characterization of our highly reflective DBRs, whose schematic view, assembled in a PhC cavity configuration, was presented in the previous section. Once that the efficient operation of our Bragg mirrors is established, the dispersion tailoring functionalities will be addressed in the following section. \\ 
The corrugation $\Gamma$ is the only quantity which is left free to vary, and we leverage it for our design purposes. The value of $\Gamma$ maps the local photonic bands diagram, and quantifies the feedback which linearly couples the two counter-propagating waves $\mathcal{A}^{\pm}$. We fix the value of $\Gamma$ in a 25 µm long central section, and we parabolically increase its value at the edges of the structure as follows:
 \begin{equation}
    \Gamma(x)=  
        \begin{cases*}
            \Gamma_0 &  for $\ x_{-}<x< x_{+}$\\
            \Gamma_0 +  \Gamma_2 x^2 &  for $\ |x|>x_{+}$
        \end{cases*} \ \ \ ,
        \label{eq:Gamma_PhC0}
 \end{equation}
\noindent where $x_{\pm}=\pm12.5$ µm, $\Gamma_0=0.1$ µm and $\Gamma_2=150$ m$^{-1}$. The longitudinal profile of $\Gamma$ is reported in Figure \ref{fig2:PhC0_design} (a). This configuration results in an adiabatic variation of the DBRs geometry, which guarantees strong (or \textit{gentle}) spatial confinement.

\noindent Figure \ref{fig2:PhC0_design} (b) presents the cavity spectrum calculated by means of a commercial FEM solver (COMSOL Multiphysics). We also show the first four modes, as predicted by the RM system Eq.\ref{eq:RM_spatial}: as can be seen, there is good quantitative agreement with the corresponding FEM solutions, which confirms the high accuracy of the RM approach \cite{Talenti_2022}. Figure \ref{fig2:PhC0_design} (c) shows the intrinsic quality factor for each corresponding mode, establishing a direct connection between the corrugation depth and the efficiency of light confinement. The curve $\Gamma(x)$, indeed, determines the shape of the potential well and the spatial profiles of the confined modes. Lower order modes, are tightly confined far from the edges of the potential well, thus exhibiting negligible radiative leakage. This explains why Q factors are typically higher for lower modal orders and larger $\lambda$.\\ 
\noindent The standard technique to reduce the radiative leakage of a 2D PhC cavity relies on fulfilling the total internal reflection (TIR) condition along the vertical (i.e. out-of-plane) direction. The latter is satisfied by minimizing the spectral components of the in-plane waves within the light cone, which is delimited by a circle of diameter $2\pi/\lambda_0$, and centered in 0, in the $k_x-k_y$ Fourier domain \cite{Akahane_2003,Akahane_2005}. Here, $\lambda_0$ is the wavelength in air, and $\textbf{k}$ is the wavevector. We illustrate these ideas in Figure \ref{fig2:PhC0_design} (d-g) for an high-order confined mode. The  2D and 1D $y$ component of the electric field intensity, $|E_y|$,  is reported in Figure \ref{fig2:PhC0_design} (d) and (e), respectively. Note that in Figure \ref{fig2:PhC0_design} (d) the scheme of the PhC cavity layout--here reported for illustrative purposes--is not in scale. We also report the corresponding Fourier transform of the field intensity in Figure \ref{fig2:PhC0_design} (f-g). We can see that the components within the light cone are negligible, thus ensuring minimal radiative losses, as by design prescriptions.\\
\begin{figure}[!b]
  \includegraphics[width=\textwidth]{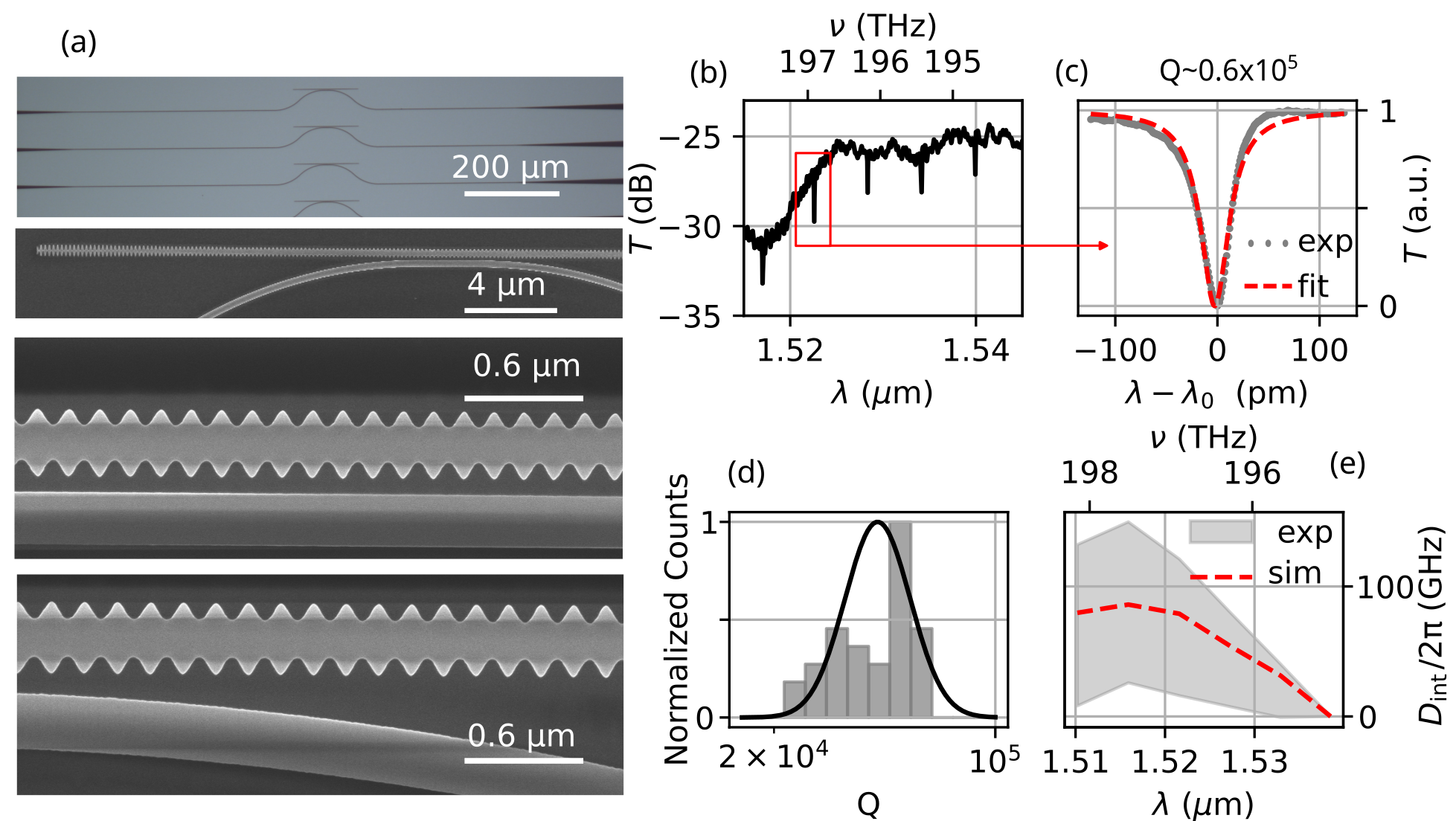}
  \caption{(a) SEM pictures of the fabricated device. (b) Transmission. (c) Resonance fitting.  (d) Loaded quality factors statistics. (e) Integrated dispersion statistics ($D_{\mathrm{int}}$).  }
  \label{fig3:PhC0_exp}
\end{figure}

In our fabricated device, light is coupled onto the chip using sub-wavelength grating couplers (GCs) \cite{Lovisolo_2025}. The fundamental transverse mode of the bus waveguide is excited, and the coupled light is guided until it reaches a region where it is evanescently coupled to the PhC cavity. We can see some snapshots of scanning electron microscope (SEM) images in Figure \ref{fig3:PhC0_exp}(a). Finding the optimal coupling conditions is not an easy task. Ideally, we target a \textit{critical} coupling, which equals the internal losses of the cavity \cite{Pasquazi_2018}. For this reason, we span over different bus waveguide to cavity coupling conditions. Among more than 50 cavities, well-defined resonances were consistently observed in a limited subset of 9 samples. In general, we observed a weak coupling and loaded quality factors of $Q_{L}\lesssim 6\times10^5$. Given that our devices are visibly under-coupled, we can estimate an intrinsic quality factor of $Q_i\gtrsim  10^5$ \cite{Vanier_2011}. Figure \ref{fig3:PhC0_exp} (b) reports resonances observed on a $\sim 30$ nm bandwidth. The sloped background originates from the bell-shaped transmission spectrum of the GCs. Figure \ref{fig3:PhC0_exp} (c) shows an example of resonance fitting, while Figure \ref{fig3:PhC0_exp} (d) presents the statistics of the quality factors.

Finally, since our goal is to establish a reliable technique for dispersion engineering, we quantified the chromatic dispersion and compared the theoretical predictions with experimental results. We introduce the \textit{integrated dispersion} $D_{\mathrm{int,m}}$ relative to the $m^\mathrm{th}$ mode as \cite{Kippenberg_2018}:
\begin{equation}
    D_{\mathrm{int,m}}=\omega_m - \omega_{\mathrm{ref}}-m\mathcal{D}_1\ \ \ \ ,
    \label{eq:Dint}
\end{equation}
\noindent where $\omega_{\mathrm{ref}}$ is a reference angular frequency and $\mathcal{D}_1$ is a reference frequency interval. In Figure \ref{fig2:PhC0_design} (e) we illustrate the statistics of the experimental (exp) $D_{\mathrm{int,m}}$ profiles, showing an optimal agreement with the theoretical predictions (sim).
\\

In summary, the fabricated AlGaAs-OI Bragg mirrors offer an efficient solution for light confinement, exhibiting a highly predictable dispersion profile which is suitable for the inverse design of the nanophotonic circuits.

\section{Dispersion tailoring by distributed feedback}
\label{sec:dispersion tailored FP-PhC}

After testing the reliable fabrication of the designed AlGaAs-OI DBRs, we have used them to customize the chromatic dispersion of the resulting cavity. In this section, we first present the results of our systematic dispersion engineering (\ref{subsec:4.1}); next, we compare these results with their experimental characterization (\ref{subsec:4.2}).

\subsection{Systematic design}
\label{subsec:4.1}

In this section, we discuss the engineering of flat dispersion profiles across various linear resonator designs, with the purpose of achieving extremely low pump-power nonlinear frequency mixing. \cite{Moille_2023,Chopin_2022}. We consider homogeneous central sections of different lengths, i.e., $L_0=0$ $\mu$m, $L_1=40$ $\mu$m, $L_2=150$ $\mu$m. From now on, we refer to the three different designs by means of the length of their homogeneous central section ($L_{0,1,2}$). As the latter increases with respect to the length of the DBRs, the reflections are more narrowly distributed on the edges of the structure. The extreme situation ($L_2$) closely mimics the case of a FP resonator, where the longitudinal reflections are strongly localized, and the wavelength dependence of the cavity length is negligible. Here, we study the phenomenological transition from a pure PhC cavity ($L_0$), towards a hybrid design which combines a free-running section with two identical and symmetric DBRs ($L_1$ and $L_2$). In each case, we aim at tailoring a prescribed dispersion profile.

Our approach closely resembles the standard inverse design (ID) techniques, which are commonly employed in nanophotonics \cite{Molesky_2018, Yang_2023}. However, a key distinction lies in the strict constraints that we impose on the \textit{n}-dimensional geometry of the PhC unit cell. Unlike conventional PhC ID methods, where the unit cell shape is subject to optimization\cite{Deng_2024}, we keep it fixed, as detailed in Section \ref{sec:intro}. Instead, our design objective is limited to the modulation of the amplitude profile, or $\Gamma(x)$, of the sinusoidal corrugation along the cavity. For further details on the design technique, we address the reader to a recent publication \cite{Talenti_2022}.\\

Let us express $\Gamma(x)$ in terms of a generalized polynomial expansion as:

\begin{subnumcases}{ \Gamma(x)=}
    0 &  for $\ x_{-}<x< x_{+}$\label{eq:FP_sec}\label{Eq:FP_section}\\
    \sum_{n=1}^N \Gamma_n |x|^n &  for $\ x_{\pm} \lessgtr x \lessgtr \pm x_{max}$\label{eq:Taper}\\
    \Gamma_{max} & for   $|x| \ge x_{max}$\label{eq:clamp}
\end{subnumcases}
\addtocounter{equation}{6}

\noindent where we fix $N=4$, and the coefficients $ \Gamma_n$, with $n=1,2,3,4$, are the subject of the optimization procedure. $N=4$ represents an optimal compromise between the required accuracy and computational cost. The central section boundaries are $x_{\pm}=\pm L_j/2$. Eventually, the parameter $\Gamma$ is clamped to a maximum value $\Gamma_{max} = \sum_{n=1}^N \Gamma_n x_{max}^n$ on the extreme edges of the structure, in order to stretch the length of the mirrors and so guarantee a stronger confinement. The object of optimization is the cost function:

\begin{equation}
    \mathrm{cost} f =\sum_m^M \dfrac{|D_{\mathrm{int},m}-\overline{D_{\mathrm{int},m}}|}{\mathcal{D}_1}
    \label{eq:cost f}
\end{equation}

\noindent where $\overline{D_{\mathrm{int},m}}$ is the target dispersion profile.

\begin{figure}[!t]
    \centering
  \includegraphics[width=0.85\textwidth]{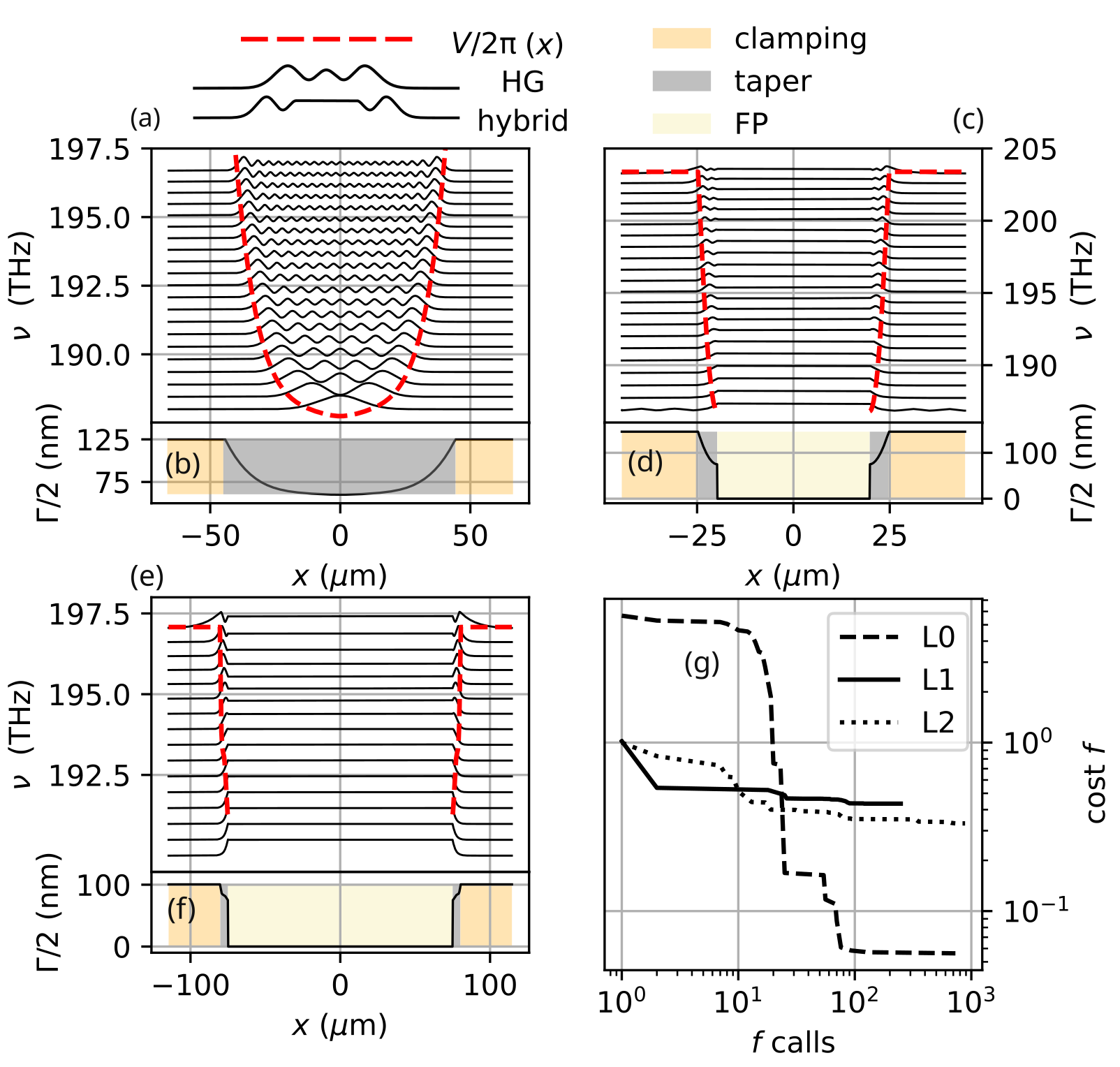}
  \caption{Spatial confinement of a flat dispersion FP-PhC cavity with a homogeneous central section of $L_0=0\ \mu$m (a), $L_1=40\ \mu$m  (c), $L_2=150\ \mu$m (e), respectively. To improve readability, only one out of every two solutions is reported in panel (e). In (b), (d) and (f) we report the corresponding sinusoidal corrugation amplitude $\Gamma(x)$. (g) Minimization of the cost function for the three different optimizations $L_{0,1,2}$.  }
  \label{fig3:ID_design}
\end{figure}

Eq.\ref{eq:cost f} was minimized by using a multi-coefficient gradient descent method (i.e. Nelder-Mead) \cite{McKinnon_1998}. We targeted a flat dispersion across $M=10$ cavity modes. The goal here is to engineer the spectral regime which is most affected by modifications to the waveguide ridge (specifically, the modes that are most sensitive to photonic bandgap tuning, representing the transition from a pure PhC to a FP resonator). In the PhC case, the chromatic dispersion of the structure is primarily governed by longitudinal variations of the waveguide geometry. In contrast, for a FP configuration, where the transverse cross-section does not change longitudinally, dispersion is governed by the wavelength dependence of the resulting effective refractive index $n_{\mathrm{eff}}(\lambda)$. \\

In Figure \ref{fig3:ID_design}(a) we illustrate the flat-dispersion PhC cavity spectrum and the confining potential well $V(x)$ resulting from the inversely designed $\Gamma (x)$ profile shown in Figure \ref{fig3:ID_design} (b). In this case, since the feedback is distributed along the entire cavity, the longitudinal dependence of the spatial confinement becomes evident, and leads to the formation of HG modes. This situation corresponds to the degenerate case $L_0=0$ µm. Next, we consider the case of a relatively small homogeneous central section $L_1=40$ µm, whose spectral confinement is shown in Figure\ref{fig3:ID_design} (c); the corresponding $\Gamma (x)$  profile is drawn in panel (d). We may notice that now the confined modes are hybridized: they exhibit a constant amplitude within the homogeneous central section, while they are HG-shaped in the tapered feedback section  $x_{\pm}\lessgtr x \lessgtr \pm x_{max}$. The  tapering domain of the optimized DBRs profile extends over just $\Delta x \equiv x_{max} -x_+\sim 5$ nm. This suffices to tailor dispersive effects over a spectral domain $\Delta \nu \sim 15$ THz, which represents the effective photonic bandgap tuning interval. Indeed the DBRs are also located at longitudinal coordinates $ |x| \ge x_{max} $ for a fixed $\Gamma_{max}$, but in this outer section they only serve to enhanced reflection, and to improve the resulting confinement. On the other hand, dispersion properties are uniquely tailored in the section where the $\Gamma(x)$ profile is tapered.\\

We marked by different colors the three characteristic sections of the $\mu$-cavity: \\
(i) the central section (\textit{FP}) is the free-running domain where the two counter-propagating waves are not linearly coupled by local reflections, being $\Gamma=0\ $nm. It is mathematically defined by Eq. \ref{Eq:FP_section}. The dynamics and the modal confinement are identical to the case of standard FP resonators. Note that this section is missing in the PhC configuration (Figure \ref{fig3:PhC0_exp} (b)).\\
(ii) In the intermediate section $x_{\pm}\lessgtr x\lessgtr \pm x_{max}$ (\textit{taper}, defined by Eq. \ref{eq:Taper}), the DBRs provide a feedback mechanism with increasing grating strength and enhanced reflectivity towards the edges of the structure. This unit tailors the dispersive effects, as it shapes the effective potential well $V(x)$ and the local coupling between the two counter-propagating waves.\\
(iii) Finally, in the most external section $ |x|>\pm x_{max}$ (\textit{clamping}, defined by Eq. \ref{eq:clamp}),) the $\Gamma(x)$ profile is kept fixed for several PhC periods, in order to improve the overall reflectivity of the Bragg reflectors.\\

Finally, in Figure \ref{fig3:ID_design} (e,f), we illustrate the ultimate design, exhibiting a central FP section of length $L_2=150$ µm. Although $L_2$ is relatively short compared to typical FP–PhC designs \cite{Yu_2019,Wildi_2023}, it is still significantly longer than the adjacent taper regions. This configuration represents the endpoint of the PhC-to-FP resonator transition that we explored in the present study. 
\\

To conclude the design description, we also report the evolution of the cost function $f$ (Eq.(\ref{eq:cost f})) vs. the function calls during the optimization of the dispersion profiles for the three different cavities $L_{0,1,2}$. As we can see in Figure \ref{fig3:ID_design} (g), the number of calls spans from a few hundreds to almost one thousand. For this reason, as anticipated in section \ref{sec:intro}, it is crucial to dispose of a fast and reliable solver. In our case, we used the 1D RM model of Eqs.(\ref{eq:RM_spatial}), which was carefully calibrated on the FEM solutions of a spare reference PhC design. In this way, we could achieve a good accuracy without loosing computational performance  \cite{Talenti_2022}. At last, our home-built solver employs $\lesssim 0.5$ secs to compute the spectrum and the corresponding $D_{\mathrm{int}}$ profile, which is used to evaluate the cost $f$, at each call. This typical CPU time should be compared with the much longer time scales (i. e. $\gtrsim$ hours), which are needed to compute the reference FEM solutions for calibration. It results that the whole optimization duration is of the order of just few minutes. \\
We observe that the cavity length $L_0$ is the most sensitive to the optimization process. This sensitivity arises from two closely related factors. First, in this configuration, which corresponds to a pure PhC cavity, the feedback is distributed across the entire confinement region. As a result, any modification directly alters the grating strength and the spatial mode profile, which in turn affects the Fourier spectrum significantly. Second, in FP-like resonators the resonances exhibit closely aligned free spectral ranges (FSRs). In this case, while our approach enables precise dispersion engineering over a broad spectral domain, it is less effective for inducing strong, localized dispersion changes.

\subsection{Experimental results}
\label{subsec:4.2}
\begin{figure}[!t]
  \includegraphics[width=\textwidth]{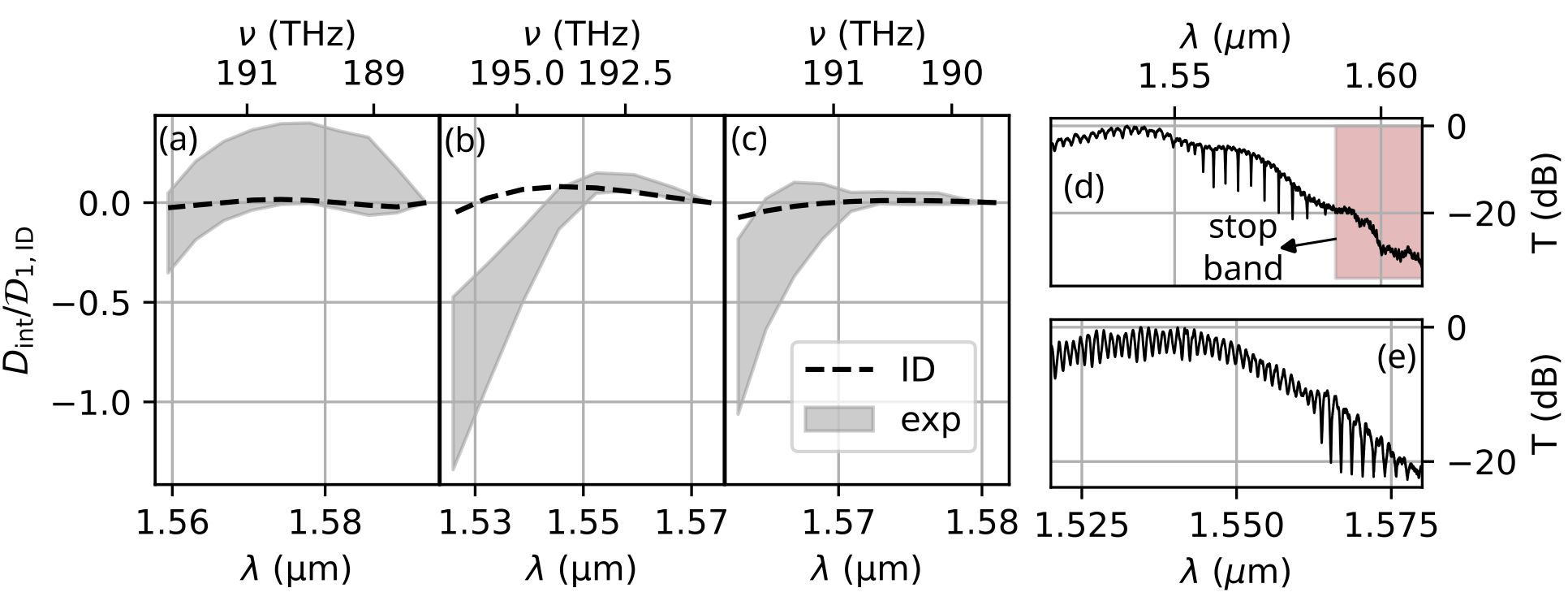}
  \caption{ Comparison between the designed and experimental $D_{\mathrm{int}}$ for the resonators (a) $L_0$, (b) $L_1$ and (c) $L_2$. Examples of optical transmission for (d) $L_0$ and (e) $L_2$. }
  \label{fig5:ID_exp}
\end{figure}

The fabricated sample contained different sets of identical cavities with different bus waveguide-to-cavity coupling conditions. We were able to observe resonances on a statistically relevant number of samples (i.e. $\geq 20$) for each cavity design $L_{0,1,2}$. The linear characterization consists in a standard transmission experiment over a $ \gtrsim 100$ nm bandwidth. We can easily measure the cavity spectrum and locate the wavelength of the resonances. The results are presented in Figure \ref{fig5:ID_exp}. In Figure \ref{fig5:ID_exp} (a,b,c) we report the theoretical dispersion profiles (black dashed line) and the corresponding confidence interval calculated from a statistical experimental data treatment (gray shadowed area). 
\\
We observe an excellent agreement between experiment and theory for the pure PhC configuration ($L_0$). In contrast, a slight deviation from theoretical predictions appears at higher frequencies for the hybrid FP–PhC designs ($L_1$ and $L_2$ ). 
We believe that the main reason for this discrepancy is due to a non-negligible group velocity mismatch $\Delta v_g \ne 0$, which arises at the boundaries between the homogeneous central section and the DBRs. This mismatch ($\Delta v_g$) not only induces scattering losses, but also a misalignment of the FSRs.
To address this problem, an intermediate section can be introduced between the FP and the tapered DBRs regions; the waveguide width could then be linearly tapered to decelerate the intra-cavity field as it approaches the DBRs from the free-running section and to minimize the discussed group velocity mismatch $\Delta v_g\rightarrow 0$ \cite{Wildi_2023}.\\
Finally, in Figure \ref{fig5:ID_exp} (d,e) we report two exemplary transmission measurements. In Figure \ref{fig5:ID_exp} (d) we highlight the emergence of the photonic bandgap, which is typical for PhC cavities.\\

\begin{figure}[!t]
  \includegraphics[width=\textwidth]{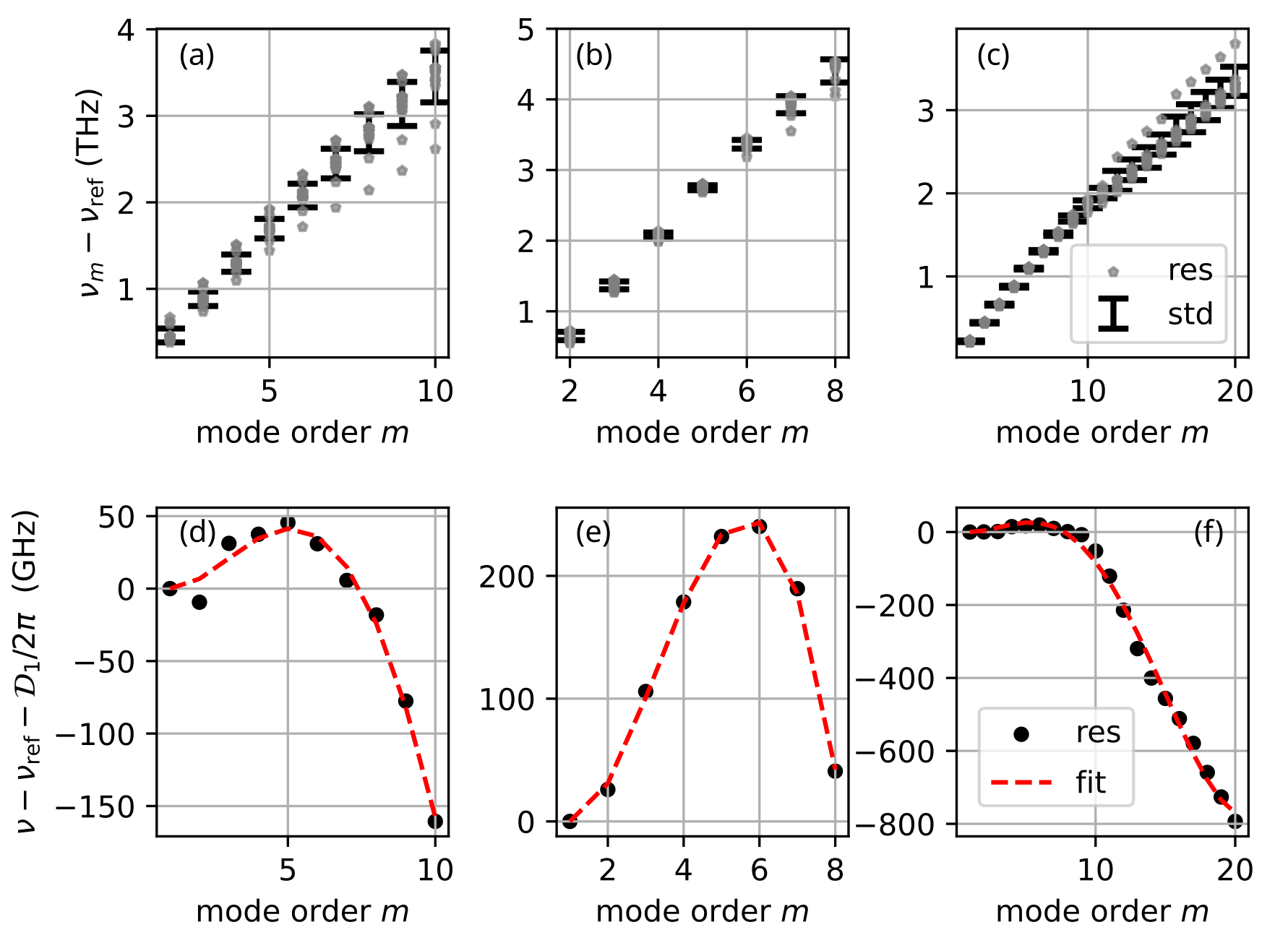}
  \caption{ Fluctuations of the measured resonances (a,b,c) and example of dispersion fitting (d,e,f) for the designs $L_0$, $L_1$ and $L_2$ (listed from left to right), respectively.}
  \label{fig6:Dint_fit_exp}
\end{figure}

In order to perform a quantitative analysis of the experimental dispersion profiles, we map the cavity resonances by expanding Eq.\ref{eq:Dint} up to the fourth order \cite{Moille_2023,Chang_2020}:
\begin{equation}
    \nu_m=\nu_{\mathrm{ref}} +\dfrac{1}{2\pi}\sum_{n=1}^4 \dfrac{\mathcal{D}_n m^2}{n!}\ \ \ .
    \label{eq:fit_formula}
\end{equation}

\noindent We fitted Eq.\ref{eq:fit_formula} for each cavity. Then, for each cavity design we regrouped the results of the fit to extract the mean value and the standard deviation (under the assumption of a Gaussian distribution).
The results of the fit are listed in Table \ref{Tab1:D12}.

\begin{table}[!h]
\centering
\caption{Fit estimation of dispersion parameters up to the fourth order ($\mathcal{D}_{1,\dots ,4}$) .}
\label{Tab1:D12}
\begin{tabular}{@{}lcccc@{}}
\toprule
\textbf{Resonator} & \textbf{$\mathcal{D}_1/2\pi$ (GHz)} & \textbf{$\mathcal{D}_2/2\pi$ (GHz)}& \textbf{$\mathcal{D}_3/2\pi$ (GHz)}& \textbf{$\mathcal{D}_4/2\pi$ (GHz)} \\
\midrule
$L_0$ & $470 \pm 90$ & $-30 \pm 60$ & $10 \pm 30$& $-3 \pm 6$ \\
$L_1$ & $310 \pm 90$ & $120 \pm 90$ & $-60 \pm 70$& $10 \pm 20$ \\
$L_2$  & $225 \pm 11$ & $2 \pm 8$ & $-3 \pm 2$  & $0.3 \pm 0.4$ \\
\bottomrule
\end{tabular}
\end{table}

Substantial deviations are observed, suggesting that fabrication tolerances may hinder the reproducibility of the present results. 
The fluctuations of the resonances for each design are shown in Figure \ref{fig6:Dint_fit_exp} (a,b,c). Here we report the measured resonances (\textit{res}) and their standard deviations w.r.t. their mean value (\textit{std}). Finally, in panel (d,e,f), we report three examples of estimated fits of Eq. \ref{eq:fit_formula} for the three dsigns $L_{0,1,2}$.\\

To summarize, we can conclude that the systematic engineering dispersion technique recently discussed \cite{Talenti_2022}--and here demonstrated for the first time--exhibits high performances for the pure PhC case, even though it slightly deviates from design prescriptions in hybrid FP-PhC configurations. Some further practical implementations could improve the current results, as the insertion of a linear tapering section to smother the group velocities mismatch between the FP and the mirroring sections \cite{Wildi_2023}, also targeting an efficient nonlinear operation \cite{Ye_2025}.

\section{Conclusions and perspectives}
\label{sec:Conclusions}
 In conclusions, we introduced a novel approach to perform a systematic design of dispersion tailored multimode linear micro-cavities. We investigated quasi-periodic PhC systems, and pointed out the strong connection between photonic bandgap tunability and dispersion engineered nanophotonics. In this context, the on-chip integration of Fabry-Pérot resonators is addressed by means of a distributed feedback mechanism. Here we studied the continuous transition from a PhC cavity into an hybrid linear resonator, comprised by two identical and symmetric distributed Bragg reflectors. The latter embeds the characteristics of both standard FP and PhC cavities, reason why it is often addressed to as FP-PhC, or, similarly, as Fabry–Pérot Bragg grating (FPBG). In our study, we implemented an optimization algorithm for targeting a prescribed dispersion profile for a FP-PhC. We made use of a reduced model, which approximates the complex 3D dynamics by means of a compact and computationally light 1D coupled waves theory \cite{Talenti_2022}. We presented an experimental demonstration of this inverse design approach based on the AlGaAs-on-insulator platform, showing a good quantitative agreement with simulations.
 \\

 We believe that our method can be easily extended to all platforms exhibiting strong index contrast between the waveguide core and the cladding. In perspective, we aim to extend the functionalities of DBRs over a broadband domain, ideally over a $\omega \rightleftarrows 2\omega$ operation band. The latter is a very challenging task at telecom wavelengths, since the photonic bandgap tunability is typically limited by the material to, at best, $\sim 50$ THz \cite{Yu_2019}. Nonetheless, we believe that a reliable design solution may be achieved, with the help of a more comprehensive theoretical investigation. \\
 Finally, similar configurations may also be adopted for $\mu$-ring devices, as recently reported for SiN, including slow-light confinement\cite{Lu_2021}, Fourier synthesis dispersion engineering \cite{Moille_2023}, or the generation of microcombs with squared frequency shaping \cite{Lucas_2023}. 
 \\To summarize, dispersion engineered nanophotonics is receiving a great deal of attention. AlGaAs-OI, as discussed in the present manuscript, represents a particularly promising solution.





\section*{Author contribution statement}
F.R. Talenti (conceptualization, theory, design, experiments, and writing) and L. Lovisolo (fabrication, design, and experiments) shared the main contribution to this work. C. Alonso-Ramos, M. Morassi, A. Lemaitre, A. Harouri, A. Gerini and L. Vivien contributed to the fabrication of the device. Z. Xiao, Z. Saleh, A. De Rossi, S. Wabnitz, and G. Leo contributed to the development of the theory and to conceiving the experiment. A. De Rossi, S. Wabnitz and L. Vivien contributed to manuscript writing. L. Vivien supervised all aspects of the work. 

\section*{Funding}

The work is partly supported by the French RENATECH network. Other funding: ANR-22-CE92-0065 (Quadcomb), EU – NRRP, NextGenerationEU (PE00000001 – program “RESTART”). L. Lovisolo acknowledges the support of French Agence-innovation defense N DGA01D22020572.

\section*{Disclosures}
The authors declare no conflicts of interest.

\bibliography{bib_file}

\end{document}